\newcommand{\CLASSINPUTtoptextmargin}{19mm}
\newcommand{\CLASSINPUTbottomtextmargin}{25mm}
\documentclass[journal, a4paper]{IEEEtran}

\IEEEoverridecommandlockouts
\normalsize
\usepackage{cite}
\ifCLASSINFOpdf
\usepackage[pdftex]{graphicx}
\else
\fi
\usepackage{amsmath}
\usepackage{array}
\ifCLASSOPTIONcompsoc
\usepackage[caption=false,font=normalsize,labelfont=sf,textfont=sf]{subfig}
\else
\usepackage[caption=false,font=footnotesize]{subfig}
\fi
\usepackage{fixltx2e}
\usepackage{stfloats}
\usepackage{url}
\usepackage{hyperref} 
\hypersetup{colorlinks = true, linkcolor=red,
	citecolor=cyan, urlcolor=gray} 

\usepackage{amssymb}
\usepackage{amsthm}
\usepackage{algpseudocode}
\usepackage{algorithm}

\usepackage{graphicx}
\usepackage{mathptmx}
\usepackage{amsmath}
\usepackage{xcolor}
\allowdisplaybreaks
\newcommand{\eqnsquez}{\!\!\!\!}

\begin{document}
\title{Eco-Vehicular Edge Networks for Connected Transportation: A Distributed Multi-Agent Reinforcement Learning Approach} 

\author{
\IEEEauthorblockN{\textbf{Md Ferdous Pervej} and \textbf{Shih-Chun Lin} }\\

\IEEEauthorblockA{Department of Electrical and Computer Engineering \\
North Carolina State University, Raleigh, NC 27695, USA\\
Email: \{mpervej, slin23\}@ncsu.edu}

\thanks{This work was supported by NC State 2019 FRPD, Cisco Systems, Inc., and North Carolina Department of Transportation (NCDOT).}
\thanks{This paper has been \textbf{accepted} for publication in \textbf{VTC2020-Fall}. Certain copyright restrictions may apply.}}

\maketitle
\thispagestyle{empty}
\pagestyle{empty}

\begin{abstract}
This paper introduces an energy-efficient, software-defined vehicular edge network for the growing intelligent connected transportation system. A joint user-centric virtual cell formation and resource allocation problem is investigated to bring eco-solutions at the edge. This joint problem aims to combat against the power-hungry edge nodes while maintaining assured reliability and data rate. More specifically, by prioritizing the downlink communication of {\em dynamic eco-routing}, highly mobile autonomous vehicles are served with multiple low-powered access points (APs) simultaneously for ubiquitous connectivity and guaranteed reliability of the network. The formulated optimization is exceptionally troublesome to solve within a polynomial time, due to its complicated combinatorial structure. 
Hence, a distributed multi-agent reinforcement learning (D-MARL) algorithm is proposed for eco-vehicular edges, where multiple agents cooperatively learn to receive the best reward.
First, the algorithm segments the centralized action space into multiple smaller groups. Based on the model-free distributed $Q$ learner, each edge agent takes its actions from the respective group. Also, in each learning state, a software-defined controller chooses the global best action from individual bests of the distributed agents. Numerical results validate that our learning solution achieves near-optimal performances within a small number of training episodes as compared with existing baselines.
\end{abstract} 
	
	
\IEEEpeerreviewmaketitle

\section{Introduction}
User-centric communication has drawn significant attention lately.
To ensure a better quality of service (QoS) and appease the data-hungry users, more networking components are being shifted towards the network edges day by day.
Vehicular networking, on the other hand, has also been evolving from the rudimentary phase to an intelligent transportation system (ITS) to guarantee public safety, lessen congestion, reduce travel time and better QoS of the vehicle users (VUs).
An advanced ITS can undoubtedly save countless lives by assuring ubiquitous connectivity and well-measured timely road hazard alerts, thus, increasing the quality of experience of the VUs.
Motivated by this, several governing bodies, such as - the United States Department of Transportation in the USA \cite{USDOT, USDOT2}, are heavily investigating more of vehicle-to-everything (V2X) communication.

Novel technologies, such as dedicated short-range communications (DSRC), cellular-V2X (C-V2X), etc., are being deemed to be coupled together \cite{ghafoor2019enabling} to induce an intelligent solution in this sector.
Note that while DSRC is an IEEE 802.11 based earlier technology, C-V2X was developed and introduced by 3GPP's Release 14 for basic safety message delivery in vehicular communication \cite{zhou2020evolutionary}.
The later releases of 3GPP focus on a more evolved system design with an advanced safety measure in addition to higher throughput, guaranteed reliability, and reduced latency.
VUs move fast on the highway causing frequent handovers for V2I communication. 
Within a very short period, the received signal strength, at the downlink VUs, can deteriorate severely in traditional network-centric communication infrastructure. 
Therefore, V2I communication is notably problematic for connected transportation.  
A potential solution to these problems should ensure universal connectivity, reliability, higher throughput, and lower latency.
In addition to that, energy-efficiency (EE) should also be considered as green communication has shown its emergency rigorously lately \cite{jain2019energy,huang2020energy}.

In the literature, there exist several works \cite{sahin2018virtual,ye2019deep,8792382,8756652,gao2019joint,guleng2020edge} addressing diverse aspects of vehicular networks. 
A downlink multicasting scenario for close-proximity vehicles was acknowledged in \cite{sahin2018virtual}. 
The author, first, created a group of vehicles as a hotspot and served the hotspot users from a single transmission point. 
Ye \textit{et al.} proposed a vehicle-to-vehicle (V2V) radio resource management (RRM) method in \cite{ye2019deep} where they used deep reinforcement learning (DRL) to scrutinize the reuse of the uplink radio resources for effective V2V communication.
A similar approach was also considered by Liang \textit{et al.}, in \cite{8792382}, for V2V applications using multi-agent RL (MARL).
Ding \textit{et al.} considered RRM for vehicular networks in \cite{8756652}.
Although the authors considered virtual cell member association and RRM in their problem formulation, they did not consider user mobility.
Note that we are considering highly mobile VUs for connected transportation, where each of the parameters needs to be chosen optimally in each transmission time interval (TTI).
Therefore, our work is fundamentally different than a single snapshot-based static user-centric approach in \cite{8756652}.

Gao \textit{et al.} proposed a joint admission control and resource management scheme, for both static and vehicular users, in \cite{gao2019joint}. 
Using the Lyapunov optimization technique, the authors showed a way to increase network throughput from the traditional network-centric approach.
Guleng \textit{et al.} considered a $Q$-learning based solution for V2V communication in \cite{guleng2020edge}.
They used a two-staged learning process for minimizing overall latency and maximizing the network throughput.
In our previous work \cite{Pervej_WSR}, we considered a throughput-optimal vehicular edge network for highway transportation, in which we achieved a maximum weighted sum rate using RL.
While the studies in \cite{ye2019deep, 8792382, 8756652, gao2019joint, guleng2020edge, Pervej_WSR} aimed at optimizing the network throughput, they did not address energy consumption issues in vehicular edge networks for smart and connected transportation.

\begin{figure}[t!] 
	\centering
	\includegraphics[width = 0.5 \textwidth, height=0.3 \textheight] {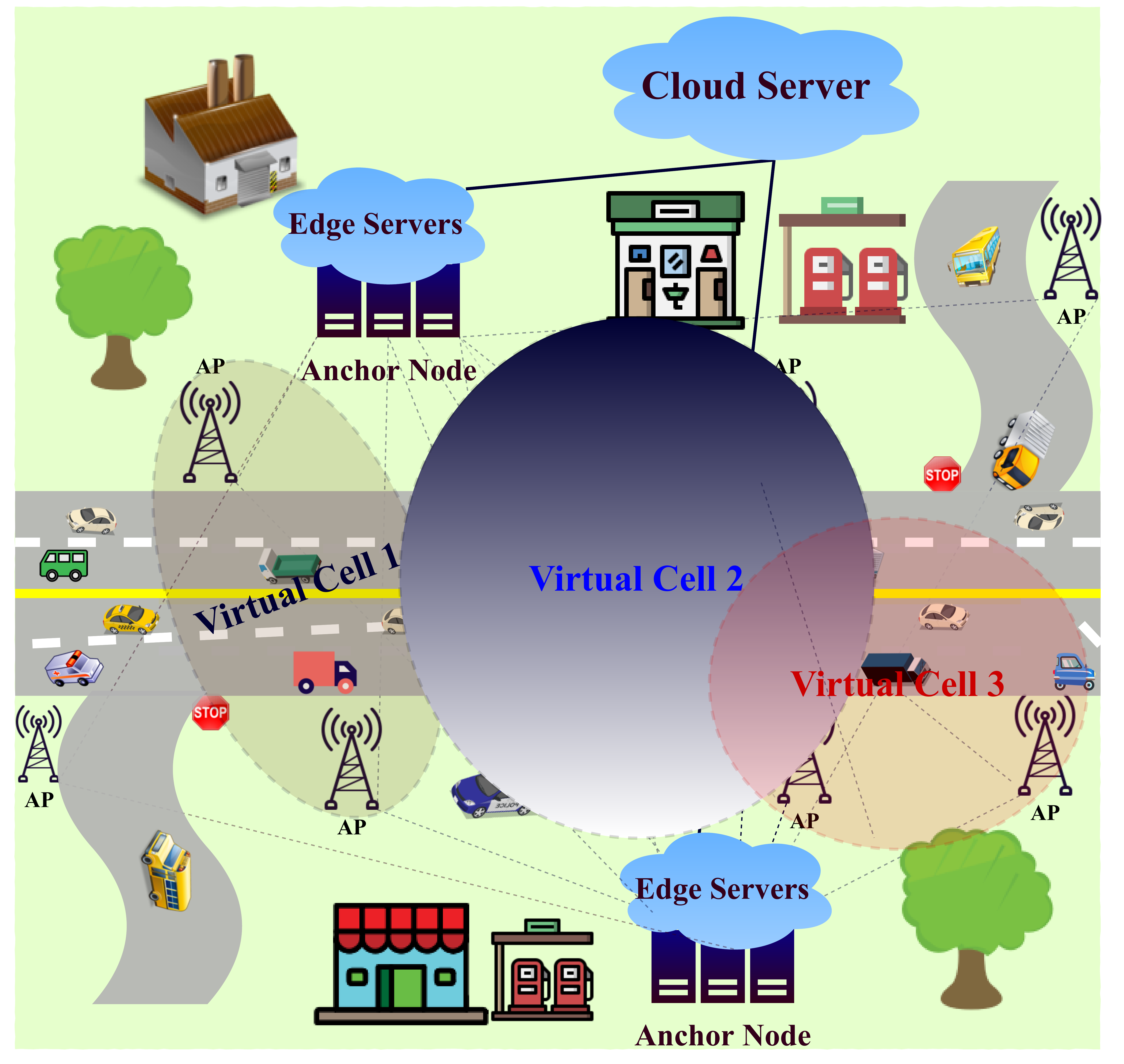} 
	\caption{Energy-efficient user-centric vehicular edge network.} 
	\label{Sys_Mod}
\end{figure}

Different from the existing studies, this paper focuses on uncovering an energy-efficient solution for user-centric and reliable vehicular edge networks in connected transportation.
Particularly, we express a joint virtual cell formation and power allocation problem for highly mobile VUs in a sophisticated SD environment.
In a freeway road environment, we delicately deploy various edge servers to obtain the users' demands by serving a VU from multiple low-powered access points (APs), as presented in Fig. \ref{Sys_Mod}.
Although such an infrastructure enhances end-to-end latency and increases the reliability of the network, system complexities also increase.
Furthermore, as multiple APs serve each VU, it is essential to optimally form virtual cells for the users and allocate the optimal transmission powers of these APs.
While our joint formulation addresses these, it is a hard combinatorial optimization problem.
Therefore, we use a model-free distributed MARL (D-MARL) solution that can effectively formulate the virtual cell and slice the resources.

\textit{To the best of our knowledge, this is the first work to consider a reliable energy-efficient user-centric software-defined vehicular edge network for connected transportation.}
The rest of the paper is organized as follows: our software-defined network and problem formulation are presented in Section \ref{system_model}. An efficient RL solution for resource slicing is presented in Section \ref{RL_Approach}.
Section \ref{Sim_Res} presents the results and findings.
Finally, Section \ref{Conclusion} concludes the paper.

\section{Software-Defined Vehicular Edge Networks}
\label{system_model}
We present our software-defined vehicular edge network model, followed by the problem formulation, in this section.

\subsection{Software-Defined System Model} 

Following the freeway case of 3GPP \cite{3gpp36_885}, in this paper, we consider a three-lane one-way road structure\footnote{We are interested in establishing a communication framework for vehicular edge networks. However, our modeling can readily be extended to a more practical environment.} as the region of interest (ROI).
Highly mobile autonomous VUs, denoted by $\mathcal{U} = \{u_1, u_2,\dots, u_U\}$, where $U \in \mathbb{Z}^+$, move on the road.
Besides, several low-powered APs, denoted by $\mathcal{A} = \{a_1, a_2,\dots, a_A\}$, where $A \in \mathbb{Z}^+$, are also deployed along the roadsides in order to maintain ubiquitous connectivity.
In addition to that, various edge servers - controlled by its respective anchor node (AN) and denoted by $b_l \in \mathcal{B}$, are deployed at a fixed known geographic position.
Each AP is physically mesh-connected to all edge servers.
Furthermore, the edge servers are connected to a centralized cloud server and have limited radio resources, denoted by $W_l$ hertz.
We consider an open-loop communication where the ANs have perfect channel state information (CSI).
Moreover, our software-defined system model is based on \cite{lin2018e2e}, where ANs can form and schedule the beamforming weights based on requirements.

By creating a virtual cell for each scheduled user, we aim to guarantee a minimum reliability threshold of the network.    
Each virtual cell contains multiple APs to serve the respective VU. 
This process of virtual cell creation is shown by the dotted ellipses in Fig. \ref{Sys_Mod}.
We denote a user and an AP by $u_i \in \mathcal{U}$ and $a_j \in \mathcal{A}$, respectively, throughout this paper.
Furthermore, the VU-AP associations are denoted by the following two indicator functions: 
\begin{equation}
\label{VU_AP_association}
\begin{aligned}
    a_i^j(t) = \begin{cases}
    1, & \text{if AP $a_j$ is associated with VU $u_i$}\\
    0, & \text{otherwise} .
    \end{cases}
\end{aligned}
\end{equation} 
\begin{equation}
\label{AP_VU_association}
\begin{aligned}
    u_j^i (t) = \begin{cases}
    1, & \text{if VU $u_i$ is associated with AP $a_j$}\\
    0, & \text{otherwise} .
\end{cases}
\end{aligned}
\end{equation} 
Therefore, $\mathcal{A}_{i}(t)$ denotes the set of APs that VU $u_i$ is connected to and $\mathcal{U}_{j}(t)$ is the virtual cell for the VU $u_i$.

\subsection{SD-V2I Communication Model} 
We consider a multiple-input-single-output communication model. 
Each VU has a single antenna, whereas each AP has $N_j$ antennas\footnote{While we consider omnidirectional antennas in this paper, the proposed framework can easily be extended with directional antennas and beam patterns to further improve SINR at vehicle receivers.}.
The wireless channel is considered to be quasi-static flat fading during a basic time block.
The channel between VU $u_i$ and the APs are denoted by $\mathbf{h}_{i}(t) = \left[\mathbf{h}_{a_1^i}^T(t), \mathbf{h}_{a_2^i}^T(t), \dots, \mathbf{h}_{a^i_A}^T(t)  \right]^T 
=  \mathbf{\mathrm{D}}_{i} (t) \rho_i(t) \zeta_{i}(t) \in \mathbb{C}^{N \times 1}$, where $\mathbf{h}_{a_j^i}(t) = [h_{ij_1}(t), h_{ij_2}(t), \dots, h_{ij_{N_j}}(t)]^T$, $\mathbf{\mathrm{D}}_{i}(t)$, $\rho_i(t)$ and $\zeta_{i}(t) \sim CN \left( \mathbf{0}, \mathbf{I}_{N}\right)$ are the channel response at a VU $u_i$ from the AP $a_j$, large scale fading, $log$-Normal shadowing and fast fading channel vectors, respectively.
Furthermore, the beamforming vector for VU $u_i$ is denoted by $\mathbf{w}_i (t) \stackrel{\Delta}{=} \left[\mathbf{w}_{a_1^i}^{T}(t), \dots, \mathbf{w}_{a_A^i}^{T}(t)  \right]^T \in \mathbb{C}^{N_i \times 1}$, where $\mathbf{w}_{a_j^i}(t) \in \mathbb{C}^{N_j \times 1}$ represents the beamforming vector of AP $a_j$ for VU $u_i$ at time $t$.
Using this beamforming vector, the transmitted signal of AP $a_j$ is denoted as $\mathbf{s}_j(t) = \sum_{i = 1}^{U} \mathbf{w}_{a_j^i}(t) x_i(t)$, where $x_i(t)$ is the unit powered signal for $u_i$ and $\mathbb{E}[x_i^H(t) x_i(t)] = 1$.
As such, at time $t$, the downlink received signal at $u_i$ is calculated as follows:
\begin{equation}
	\label{rx_signal_at_UE}
	\begin{aligned}
	\eqnsquez y_{i}(t) &= \sum_{a_j \in \mathcal{A}} \mathbf{h}_{a_j^i} ^H (t) \mathbf{s}_j(t)  + \eta_i(t) \\
	&= \mathbf{h}_{i}^H(t) \mathbf{w}_i(t) x_i(t) + \sum_{u_{i'} \in \mathcal{U}\backslash u_i} \mathbf{h}_{i}^H(t) \mathbf{w}_{i'}(t) x_{i'}(t) + \eta_i(t), 
	\end{aligned}
\end{equation} 
where $\eta_i(t)$ is the received noise at time $t$. Besides, $\eta_i(t)$ is circularly symmetric complex Gaussian distributed with zero mean and $\sigma^2$ variance. 

\subsection{User-Centric Dynamic Cell Formation} 
\label{UE_Centric_Cell}

The vehicular edge network is considered to operate in time division duplex mode.
Thus, we calculate the achievable rate for VU $u_i$, at time $t$, as follows: 
\begin{equation}
\label{ergodic_data_rate}
    R_i^t \left(\mathbf{W(t)}\right) = \left(1-\kappa \right)\log_2 \left(1 + \frac{\left| \mathbf{h}_{i}^H(t) \mathbf{w}_{i}(t) \right|^2}{\sigma^2 + \sum_{u_{i'} \in \mathcal{U} \backslash u_i}  \left|\mathbf{h}_{i}^H(t) \mathbf{w}_{i'}(t) \right|^2} \right),
\end{equation}
where $\kappa$ is spectral efficiency loss due to signaling at the APs and $\left| \mathbf{h}_{i}^H(t) \mathbf{w}_{i}(t) \right|^2 / \left(\sigma^2 + \sum_{u_{i'} \in \mathcal{U} \backslash u_i}  \left|\mathbf{h}_{i}^H(t) \mathbf{w}_{i'}(t) \right|^2 \right) = \gamma_{i}^t\left(\mathbf{W}(t)\right) $ is the SINR. 
Moreover, as multiple APs are scheduled to transmit to $u_i$, the backhaul link consumption by the VU is carefully calculated as follows\cite{6831362}:
\begin{equation}
\label{backhaul_consump_u_i}
    \mathrm{C}_i(t) = \left \Vert \left[ \left \Vert \mathbf{w}_{a_1^i}(t) \right\Vert_2, \dots, \left \Vert \mathbf{w}_{a_A^i}(t) \right\Vert_2\right]  \right \Vert_0 R_i^t \left(\mathbf{W}(t)\right),
\end{equation}
where $\left \Vert \cdot \right \Vert_0$ denotes the total number of nonzero elements in a vector. 
This is commonly known as the $l_0$-norm.
If a user is scheduled in a transmission time slot $t$, the precoding vectors from all of the APs for that VU, i.e., $\mathbf{w}_i(t)$, is nonzero leading to a nonzero achievable data rate.

Note that we presume to serve all active users, in a transmission time slot, by forming virtual cells for each user and dynamically allocating transmission power of the APs.
As such, we intend to find optimal user-centric cell formation and beamforming weights calculation - for the APs, in our objective function.
The first question that we try to answer is - \textit{what is the maximum throughput in our SD controlled highly mobile vehicular network?}
A naive approach would be serving a user from as many APs as possible with the maximum transmission powers of the APs.
However, this will bring down the user fairness and EE whatsoever.
Therefore, it is essential to justify the user data rate with EE. 
To avoid cross-domain nomenclature, let us define what we refer to as the EE. \textit{The fraction of the total user sum rate to the total power consumption of the network is defined as EE}.
At a given time slot $t$, we calculate EE as follows:
\begin{equation}
\label{EE}
EE(t) = \frac{\sum_{u_i\in \mathcal{U}} \mathrm{C}_i(t)}{\sum_{a_j \in \mathcal{A}} \sum_{u_i \in \mathcal{U}} \left \Vert \mathbf{w}_{{a}_j^i}(t)\right \Vert_2^2},
\end{equation} 
where $\mathrm{C}_i(t)$ is calculated in Equation (\ref{backhaul_consump_u_i}).

Therefore, in this paper, we address the following question: \textit{what are the user-centric associations and power allocations that guarantee reliability, programmability, and EE of the entire network? }
To this end, we formulate a joint optimization problem as follows:
\begin{subequations}
	\label{Original_Problem}
	\begin{align}
	& \textbf{Find:}  && a_i^j (t),  u_j^i(t),  \mathbf{w}_{a_j^i}(t), ~~ \forall i \in \mathcal{U}, j \in \mathcal{A}  \nonumber \\ 
	\label{OP1}& \textbf{Maximize} &&  EE(t) \\
	\label{Cons_1} & \textbf{Subject to} && 1 \leq |\mathcal{A}_{ij}(t)| \leq A,~~ \forall i \in \mathcal{U}\\
	\label{Cons_3} &~&& \gamma_{i}^t\left(\mathbf{W}(t)\right) \geq \gamma_i^{min}, ~~\forall i \in \mathcal{U}\\
	\label{Cons_4} &~&& \sum_{i = 1}^U \left \Vert \mathbf{w}_{a_j^i}(t) \right \Vert_2^2 \leq P_j^{max},~~ \forall j \in \mathcal{A}\\
	\label{Cons_9} &~&& a_i^j(t) \in \{0,1\}, u_j^i(t) \in \{0, 1\},
	\end{align}
\end{subequations}
where $\gamma_i^{min}$ is the minimum SINR requirement for our reliable communication.
The reliability constraint is reflected in equation (\ref{Cons_3}).
$P_j^{max}$ is the maximum allowable transmit power of AP $a_j$ which is controlled via equation (\ref{Cons_4}).
Equation (\ref{Cons_1}) is taken to ensure each virtual cell contains more than one APs. 
Moreover, Equation (\ref{Cons_9}) indicates the feasible solution space.

Note that the $l_0$ norm restricts using the gradient-based solution. Besides, the formulated problem is a hard-combinatorial problem, which is extremely difficult to solve within a short period.
Moreover, for each of the AP, at each time slot $t$, there are $2^U-1$ possible combinations only for the possible VU-AP associations.
For each of these associations, the AP, furthermore, needs to choose the optimal power level for the scheduled users.
Note that, in this paper, instead of a continuous power level, we divide the AP's transmission power level into multiple discrete levels.
As our SD controlled ANs know the perfect CSI, we model the beamforming vector as follows:
\begin{equation}
\label{beam_power}
    \mathbf{w}_{a_j^i}(t) = \frac{\mathbf{h}_{a_j^i}(t)}{\left \Vert \mathbf{h}_{a_j^i}(t) \right \Vert_2} \times \sqrt{P_{a_j^i}(t)}, 
\end{equation} 
where $\mathbf{h}_{a_j^i}(t)$ is the wireless channel information from AP $a_j$ to VU $u_i$ and $P_{a_j^i}(t)$ is the allocated transmission power of AP $a_j$ to transmit to VU $u_i$.
If a centralized decision has to be taken, the centralized agent needs to make a central decision for all of the AP-VU associations and their power level selections.
In that case, the size of the action space is $\mathcal{O}\left((2^U-1)^A  \times K^{UA} \right)$, where $K$ is the total discrete power levels.
Thus, traditional optimization methods may take an enormous amount of time to solve such an intricate problem.
As such, we use a model-free $Q$-learning approach to solve the optimization problem efficiently in the next section.

\section{Energy-Efficient Resource Slicing at Edges: A Reinforcement Learning Approach}
\label{RL_Approach}

As we assume the CSI is known, our state-space contains all CSIs - denoted by $\mathbf{H}^t$, the locations of the VUs - denoted by $\mathbf{X}_i^t$, and the locations of the APs - denoted by $\mathbf{X}_j^t$.
Therefore, we denote the state-space by $\mathbf{s}_t = \left\{ \mathbf{X}_i^t, \mathbf{X}_j^t, \mathbf{H}^t \right\}$.
On the other hand, the action space contains the VU-AP association and beamforming vectors for the chosen association. 
The action space is, thus, a two-step process. 
First, the RL agent needs to choose a possible association.
After that, it designs the beamforming vectors.
We express the action space as $\mathrm{\mathbf{a}}_t = \left\{ a_j^i(t) ~\forall i \in \mathcal{U} , P_{a_j^i}(t) ~\forall i \in \mathcal{U}~\& ~j \in \mathcal{A} \right\}$. 
Moreover, we have taken the EE in Equation (\ref{EE}) as the reward function of the RL agent.
However, to ensure fairness among users achievable rate and reliable communication, at each time slot $t$, we have employed the following restriction:
\begin{equation}
\label{RL_Reward}
    r_t = \begin{cases}
    EE(t), & \text{if } \gamma_{i}^t\left(\mathbf{W}(t) \right) > \gamma_i^{min} ~\forall u_i \in \mathcal{U}\\
    0, & \text{otherwise}.
\end{cases} 
\end{equation}

\subsection{Single Agent Reinforcement Learning (SARL)} 
Taking the state-space and action-space into account, $Q$-learning based RL framework can effectively solve hard optimization problems.
Note that it is a model-free learning \cite{watkins1992q,lin2016qos} process, where in each state $\mathbf{s}_t$, the agent takes an action $\mathbf{a}_t$, gets a reward $r_{t}$ for the chosen action and the environment transits to the next state $\mathbf{s}_{t+1}$.
The governing equation of Q-learning is shown in the following: 
\begin{equation}
\label{q_table}
    Q(s_t,a_t) \!\!\leftarrow\!\! (1-\alpha) Q(s_t, a_t) + \alpha \left(r_t + \gamma \underset{\textbf{a}}{\text{ max }} Q(s_{t+1},\textbf{a})\right)\!,
\end{equation} 
where $\alpha$ and $\gamma$ are learning rate and discount factor, respectively.

Although SARL is a good baseline scheme, if the number of states and actions is too large, it may become impracticable to handle.
For example, if $U=A=3$ and $K=4$, then the baseline centralized SARL has an action space\footnote{Note that this contains the total action space. The number of valid actions will be lesser than this due to the maximum allowable transmission power constraint of the AP.} of order $\mathcal{O}\left((2^U-1)^A  \times K^{UA} \right) = 89915392$.
This is commonly known as the curse of dimensionality.
Please note that, in a learning environment, if the number of states is infinite, usually we approximate the state using some means of approximation \cite{sutton2018reinforcement}, such as linear approximation, neural network, etc.
However, the number of action remains unchanged.
For each approximated state, the learning agent still must take action from the original large action-space \cite{sutton2018reinforcement}.
As an alternative, we propose a D-MARL solution in what follows.

\subsection{Distributed Multi-Agent RL (D-MARL)} 
The fundamental assumption in traditional MARL is allowing multiple agents to take their independent decisions so that we have a shrunk action-space for each agent. 
Then, the idea is to collaboratively learn a joint action set such that optimal network performance can be achieved. 
Although the action space for each agent is small compared to that of the centralized SARL\footnote{For the same example of the centralized SARL one, if we consider each AP as an independent agent for the MARL scheme, the order of the action space for an agent is $\mathcal{O}\left((2^U-1) \times K^U \right) = 448$.}, whether MARL will accomplish the optimal solution within reasonable training episodes is uncertain as all learning agents need to reach to a consensus cooperatively for maximizing the reward.
Therefore, we have used the concept of MARL.
Yet, instead of letting multiple-agents taking independent actions from a shrunk action space, we have used a distributed learning process where each agent takes decisions from a segmented original SARL's action space.
In other words, the original SARL's action space is subdivided into multiple groups. 
Each agent takes its decision from an assigned smaller group. 

If there are $N$ such agents, then the dimension of the $Q$-table of such an agent is $\mathbb{R}^{S \times A/N}$, where $S$ and $A$ represents the size of the state space and action space, respectively.
Therefore, the order of the action space of each agent is of $\mathcal{O}\left(\Phi \right)$, where $\Phi = \left[ (2^U-1)^A  \times K^{UA} \right]/N$.
Furthermore, let us assume there is a centralized vector - denoted by $\textbf{Q}_\text{central} \in \mathbb{R}^{S}$, that stores the global best action at every state.
We update this global best action using the following equation: 
\begin{equation} 
\label{Q_central_update}
    \textbf{Q}_\text{central} [\textbf{a}_t] \leftarrow \begin{cases}
    \textbf{Q}_\text{central} [\textbf{a}_t], ~\text{if any } r_t [\textbf{a}_t] > r_t [\textbf{a}_t^{\text{old}}], \forall \text{agents} \in N\\
    \textbf{Q}_\text{central} [\textbf{a}_t^{\text{old}}], \text{ otherwise}. 
    \end{cases} 
\end{equation} 
Therefore, our proposed D-MARL solution distributively learns to take the optimal central action quickly.
On the other hand, traditional MARL \cite{8664589, liu2019trajectory} may not achieve the optimal solution, as independent agents take autonomous actions in a shrunk action space, within a reasonable time.
The joint actions of these agents may not be centrally optimal and lead to a sub-optimal solution.
Algorithm \ref{SARL_MARL} summarizes our proposed D-MARL solution.
\begin{algorithm} 
\caption{Distributed Multi-Agent RL (D-MARL)}
	\begin{algorithmic} [1]
		\State \textbf{Initialize}: Total number of agents, $N$  \Comment{Choose $N$ so that $Actions/N \in \mathbb{Z}^+$} 
		\State Generate random $\mathbf{Q}_l(\mathbf{s}_t, \mathbf{a}_t)$ tables, where $l \in N$.
		\State Generate $\textbf{Q}_\text{central} \in \mathbb{R}^S$ randomly
		\For {each episode}
		\State Initiate the environment, generate $	\mathbf{s}_t = \left\{ \mathbf{X}_i^t, \mathbf{X}_j^t, \mathbf{H}^t \right\}$
		\While {not terminated}
		\For {each $l \in N$}
		\State Observe the environment; choose $\textbf{a}_t$, based on the observation, following $\epsilon$-greedy policy; receive reward $r_t$; update its $Q$-table using equation (\ref{q_table})
		\If {$r_t >$ reward using $\textbf{Q}_\text{central}[\textbf{a}_t]$ }
		\State update $\textbf{Q}_\text{central}$ using equation (\ref{Q_central_update})
		\EndIf
		\EndFor
		\State $\mathbf{s}_t \leftarrow \mathbf{s}_{t+1}$
		\EndWhile \Comment {If $\mathbf{s}_t$ is the terminal state}
		\EndFor 
	\end{algorithmic} \label{SARL_MARL}
\end{algorithm}

\section{Performance Evaluation} 
\label{Sim_Res}

\begin{figure} 
    \centering
    \includegraphics[width = 0.5 \textwidth]{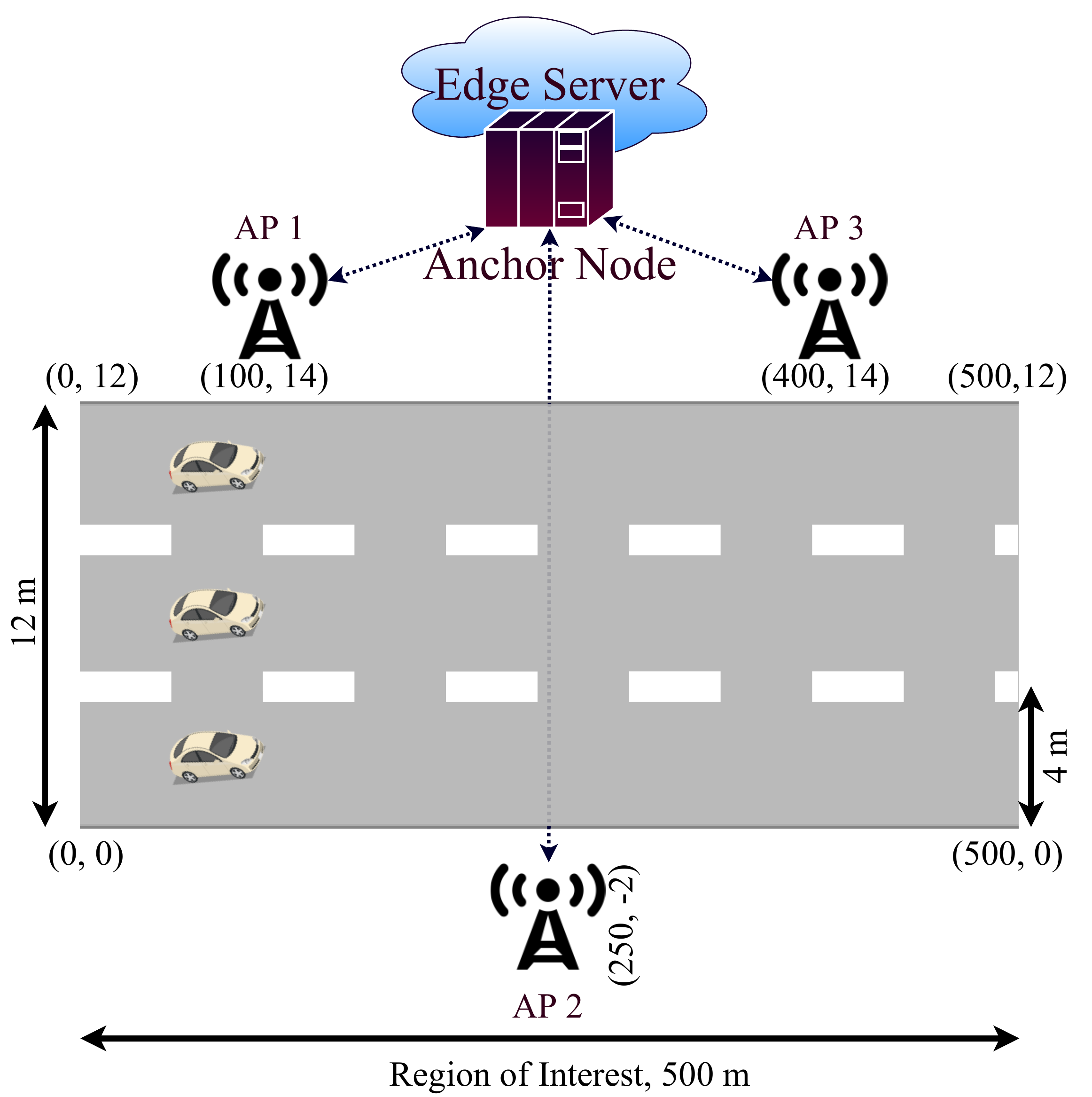} 
    \caption{Simulation environment} 
    \label{Sim_Set}
\end{figure}

We consider ROI = $500$ m, VU velocity = $140$ km/h, $|\mathcal{U}|=|\mathcal{A}|=3$, $|\mathcal{B}|$ = $1$, $8$ antenna per AP, noise power = $-114$ dBm/Hz, $\kappa=0.1$, and TTI = $100$ milliseconds.
The channels, path loss, and shadowing are modeled following \cite{3gpp36_885}.
For the ease of simulation, we consider a full buffer network model where all APs serve all VUs simultaneously.
We consider the following association rule:
\begin{equation}
\label{association_rule}
\begin{aligned}
    a_j^i(t) = \begin{cases}
    1, & \text{if $u_i$ is in the coverage region of AP $a_j$}\\
    0, & \text{otherwise}.
    \end{cases}
\end{aligned}
\end{equation}
Note that our proposed problem solution can work in other scheduling algorithms as well. 
While the VUs are dropped uniformly in each lane, the APs are placed $150$ meters apart fixed locations.
For a tractable state space, we have considered that, at a given time step, all VUs are in the same $x$ locations - while they have different $y$ locations.
The simulation setup is presented in Fig. \ref{Sim_Set}.
Here we stress out that while we consider a restricted simplified version of the environment for our simulation, our proposed algorithm shall achieve similar performances in a more comprehensive platform.
To show the effectiveness of our proposed solution, we compare its performances to the optimal benchmark performance. 
We apply certain restrictions only due to limited computation resource availability.

\begin{table*}
\caption{Performance Comparisons: $\gamma_i^{min}=10~dB$, AP Coverage Radius $= 250~m$} 
	\centering
	\begin{tabular}{|c|c|c|c|c|c|}
		\hline 
		\textbf{Scheme Name} & \textbf{Training Episodes} & \textbf{Test Episodes} & \textbf{Average EE} [bits/Hz/J] & \textbf{Deviation from Benchmark}\\ \hline 
		Brute Force (Benchmark) & N/A & $250$ & $2.319512196$ & $0 \%$ (Benchmark) \\ \hline 
		D-MARL (Proposed) & $25000$ & $250$ &  $2.319498972$ & $0.0005\%$ \\ \hline 
		SARL \cite{lin2016qos} & $100000$ & $250$ & $2.319509675$ & $0.0001 \%$\ \\ \hline 
		MARL \cite{8664589} & $100000$ & $250$ & $2.289516849$ & $1.2932 \% $ \\ \hline 
		Equal Power & N/A & $250$ & $0.002361365$ & $99.8982 \%$ \\ \hline 
		Random Power & N/A & $250$ & $0.038780234$ & $98.3281 \%$ \\ \hline 
	\end{tabular}
	\label{Performance_Compare}
\end{table*}

\subsection{Average Energy-Efficiency Comparisons}
To this end, we show the effectiveness of the proposed D-MARL algorithm. 
Specifically, we compare our design with the following schemes: 
\begin{itemize} 
\item \textit{Brute Force (Benchmark)}: This is the optimal solution. 
	In this case, at each state, we need to search for the optimal action that provides the maximum reward.
\item \textit{SARL} \cite{lin2016qos}: This is the baseline RL scheme. 
	We adopted the learning process in \cite{lin2016qos} for this case.
	Note that this is a centralized learning process. In this case, the RL agent has a large action-space. Such a centralized learning agent shall take a sufficiently large number of training episodes to reach the optimal solution. 
    \item \textit{MARL} \cite{8664589}: We have used the novel cooperative MARL learning process proposed by Yao \textit{et al.} in \cite{8664589}. 
    Note that a similar approach is also considered by Liu \textit{et al.} in \cite{liu2019trajectory}.
    Here we stress out that while there exist other MARL models, we used recently proposed collaborative learning approaches of \cite{8664589} and \cite{liu2019trajectory} for our performance comparison.
    We intend to validate that the proposed D-MARL algorithm delivers near-optimal performances within a nominal number of training episodes.
    \item \textit{Equal Power Allocation}: In this case, we have assumed that the AP divides its transmission power equally to serve the VUs. Essentially, this is the centralized case where the central power allocation decision is chosen in such a way that each AP transmits to its scheduled users using equal power.
    \item \textit{Random Power Allocation}: We have assumed that the AP chooses random transmission power from the discrete power level to serve a VU. This is also a centralized case in which, at each state and time slot, we choose a random central decision from the possible centralized action space.
\end{itemize}
We use each AP as an independent agent for the MARL algorithm.
Therefore, there are three agents for MARL \cite{8664589}, where each AP takes its association and power allocation decision independently. 
For our proposed D-MARL algorithm, we have used four agents. 
The SARL and MARL models are trained on $1\times 10^5$ episodes, whereas the D-MARL model is trained only on $25 \times 10^3$ episodes.
We have taken $\gamma = 0.8$.  
The values of both $\epsilon$ and $\alpha$ are decayed linearly from $1$ to $0.01$ in each episode.

From $250$ test episodes, the performance comparisons of our proposed algorithm with other schemes are listed in Table \ref{Performance_Compare}.
Note that we have taken $\gamma_i^{min} = 10$ dB and AP coverage radius $=250$ m for this comparison.
Clearly, machine learning solutions achieve much higher performances than two baseline schemes (equal power allocation and random power allocation). 
Furthermore, thanks to RL, the centralized baseline SARL solution and the proposed D-MARL solution deliver nearly identical performance to that of the brute force optimal performance. 
The agents learn to take optimal actions from the training episodes and deliver a near-optimal performance.
The performance of the MARL \cite{8664589} is also very close to this optimal solution. 
However, recall that SARL and MARL models are trained on four times the training episodes used in our proposed D-MARL algorithm.
These results also substantiate that a distributed learning choice on the centralized action-space is beneficial over learning on a shrunk action-space as in \cite{8664589,liu2019trajectory}.
Furthermore, our proposed D-MARL solution achieves $\approx 29$ dB and $\approx 18$ dB performance gain over equal power allocation and random power allocation schemes, respectively.	
\begin{figure}
	\centering
	\includegraphics[width = 0.5 \textwidth]{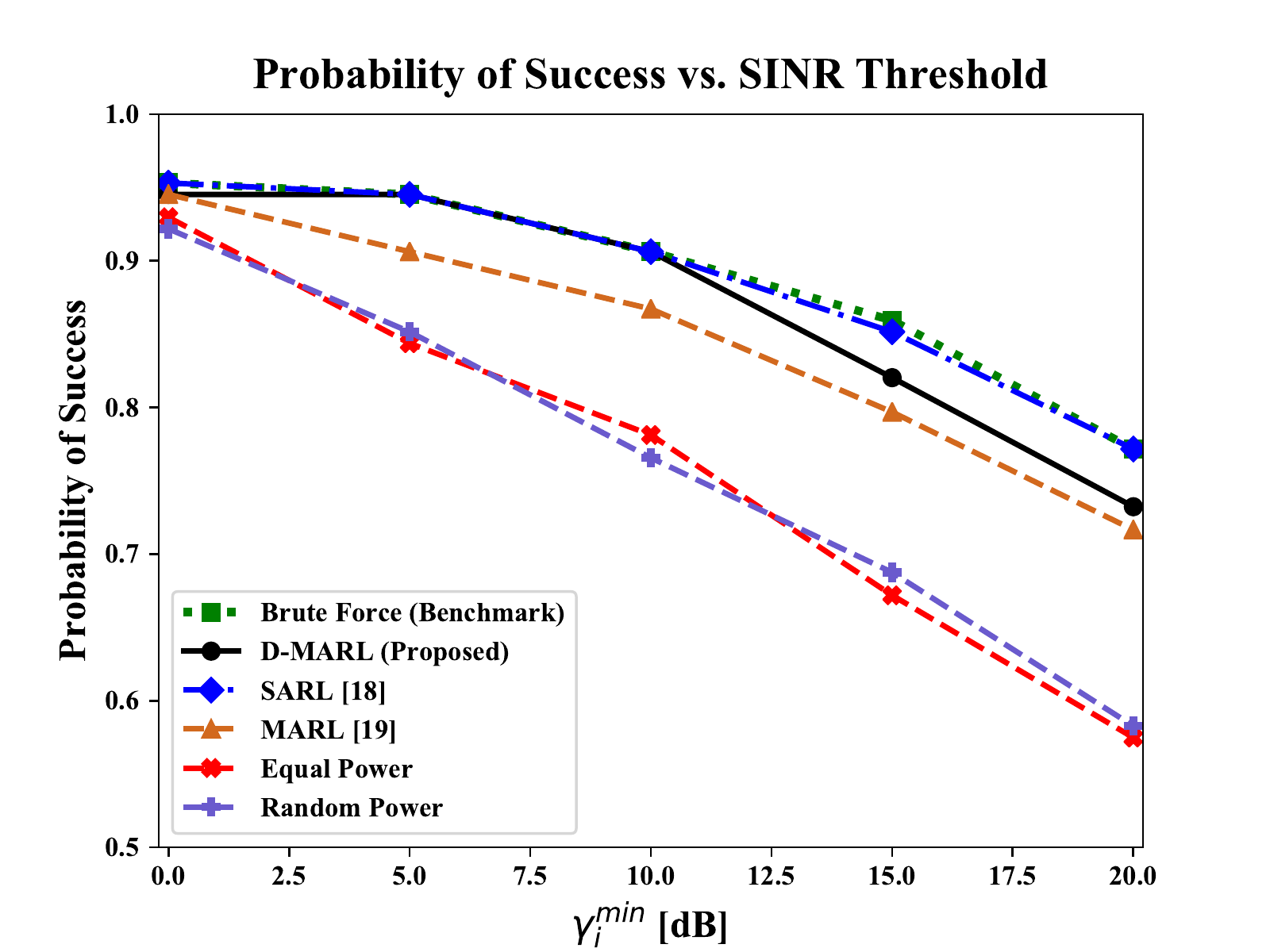} 
	\caption{Probability of success for different SINR threshold when AP coverage radius is $250$ m} 
	\label{Pro_suc}
\end{figure}

\begin{figure}
	\centering
	\includegraphics[width = 0.5 \textwidth]{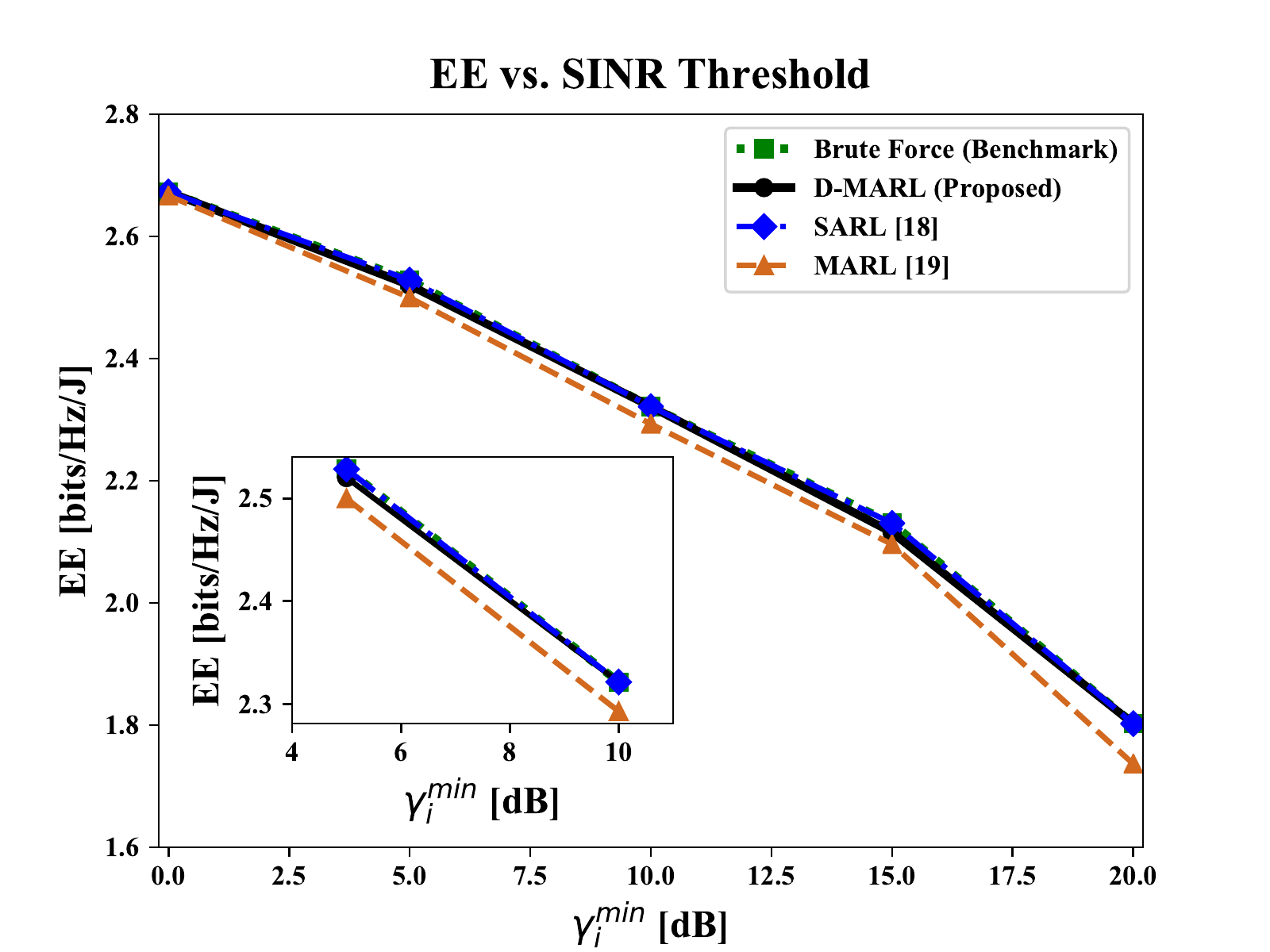} 
	\caption{Average EE for different SINR threshold when AP coverage radius is 250 m} 
	\label{EE_Vs_Schemes}
\end{figure}

\subsection{Impact of the Reliability Constraint}
The reliability constraint has a significant impact on the overall network performance.
If we increase the reliability constraint, $\gamma_i^{min}$, we force the RL agents to find optimal solutions that maximize the EE without violating the reliability constraint.
Therefore, as this constraint increases, the number of total failed events also increases.
First, let us calculate the success probability of achieving the reliability constraint as follows:
\begin{equation}  
\label{pro_suc_calculation}
	 \mathrm{Pr} ~\{\text{success}\} = 1 - \frac{1}{T}\sum_{t \in T} \mathrm{I}_f(t),
\end{equation}
where $T$ is the total number of time steps and $\mathrm{I}_f(t)$ is an indicator function for the event that $\gamma_{i}^t \left(\mathbf{W}\right) < \gamma_i^{min} $ for any of the $u_i \in \mathcal{U}$.
The probability of delivering the minimum required SINR is shown in Fig. \ref{Pro_suc}.
The RL algorithms perform better than the baseline schemes. 
Furthermore, as $\gamma_i^{min}$ increases, the successful transmission events get decayed. 
Note that our proposed D-MARL can deliver near-optimal success probability with these varying reliability requirements. On the other hand, the performance gap between MARL \cite{8664589} and D-MARL is quite evident from this result. Moreover, increasing the reliability constraint may necessitate the APs to transmit to the VUs with more power - so that it can attain the SINR threshold. However, this downgrades the EE. Our simulation results in Fig. \ref{EE_Vs_Schemes} also reflects this.
As the performances of the two baseline schemes (equal power allocation and random power allocation) are very poor compared to the RL schemes, hereinafter, we will only compare the performance of our proposed algorithm with the brute force (benchmark) and other two RL schemes.

\begin{figure}
	\centering
	\includegraphics[width = 0.5 \textwidth]{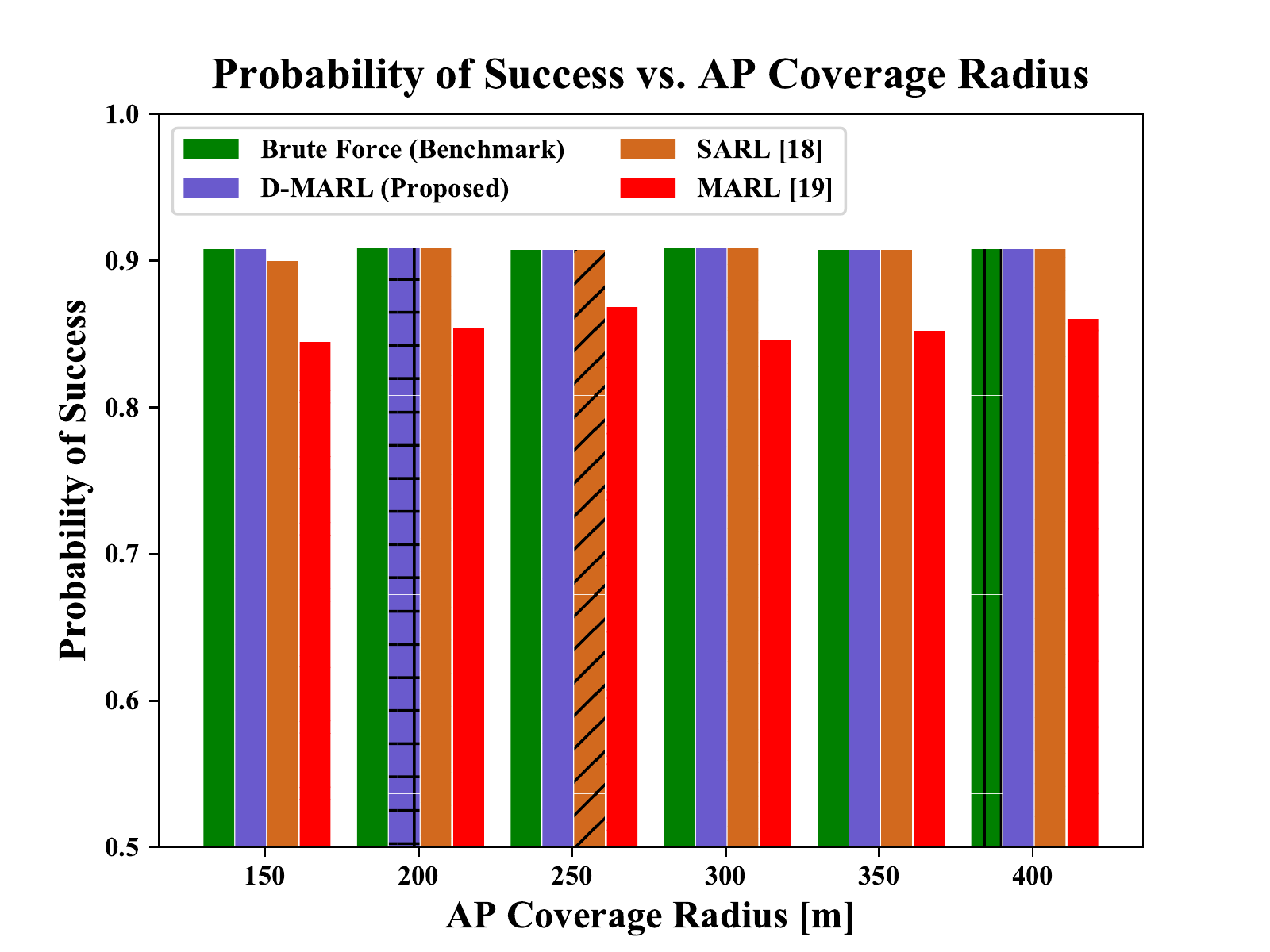} 
	\caption{Probability of success for different AP coverage radius when SINR threshold is $10$ dB} 
	\label{Pro_Suc_Radii}
\end{figure}

\begin{figure}
	\centering
	\includegraphics[width = 0.5 \textwidth]{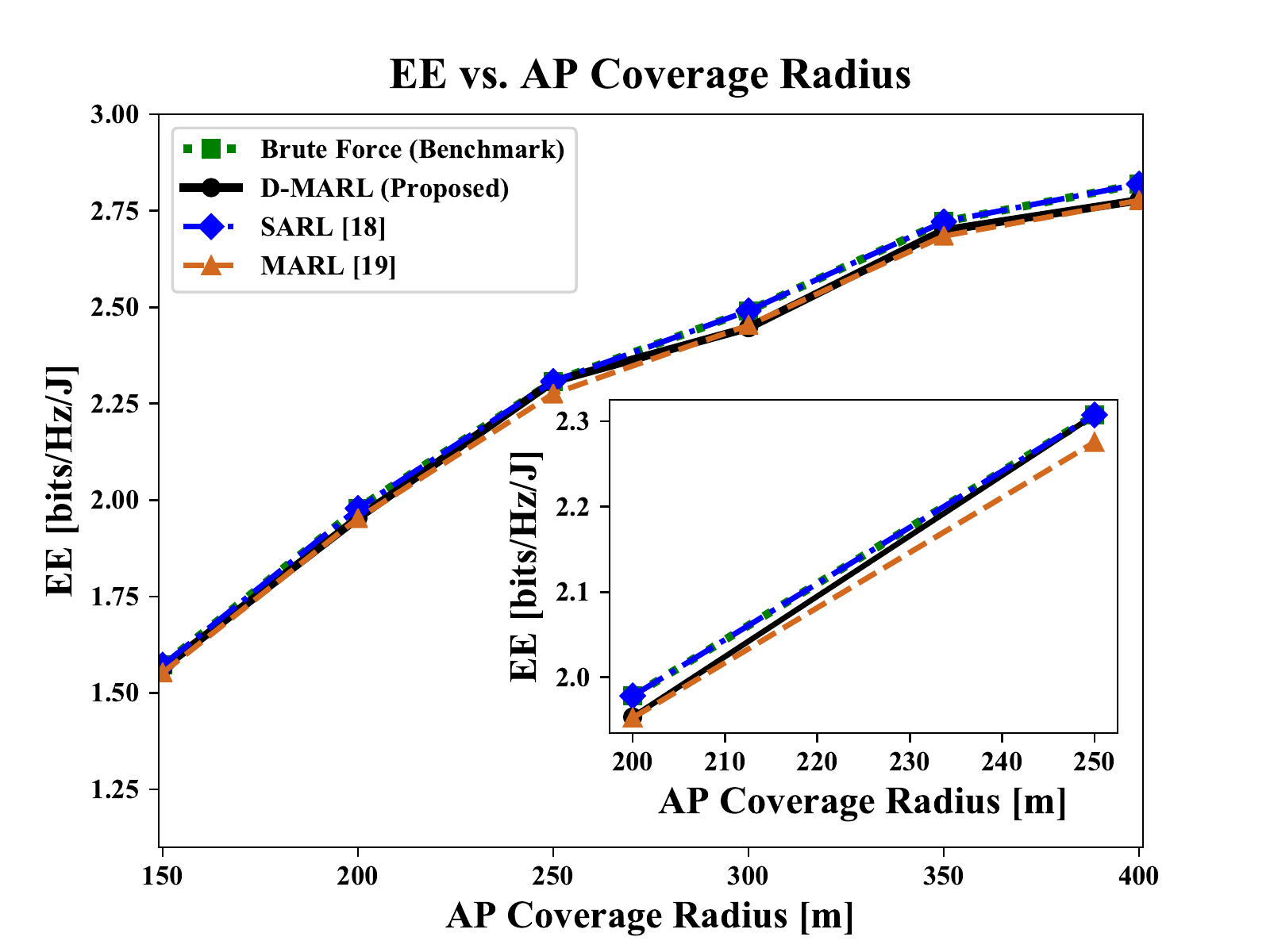} 
	\caption{Average EE for different AP coverage radius when SINR threshold is $10$ dB} 
	\label{EE_Radii}
\end{figure}

\subsection{Impact of the  Coverage Radius}
Now, we analyze the impact of the coverage radius of the APs by keeping the reliability constraint fixed and varying the coverage radius.
Note that as the reliability constraint is fixed, the probability of success, as shown in Equation (\ref{pro_suc_calculation}), should not fluctuate rapidly if the coverage radius is varied. 
This is also reflected in Fig. \ref{Pro_Suc_Radii}.
Besides, as the coverage radius of the AP increases, more VUs can be served by each of the APs.
Although the SINR constraint is fixed, recall that from our association rule in Equation (\ref{association_rule}) and rate calculation in Equation (\ref{backhaul_consump_u_i}), it is quite clear that increasing the coverage radius will increase the total number of links for a VU. 
This will, therefore, improve the user sum rate. 
On the other hand, if the VU is far away from an AP, the AP might need to transmit to it with more power.
However, the RL agents will find optimal power allocations to increase the user sum-rate that ultimately increases the EE in Equation (\ref{EE}).
This trend is also reflected in our simulation results in Fig. \ref{EE_Radii}.
As the coverage radius increases, the D-MARL algorithm finds optimal associations and power allocations, leading to an improved EE of the network.

\subsection{User Fairness}
\begin{figure} 
\centering
	\includegraphics[width = 0.5 \textwidth] {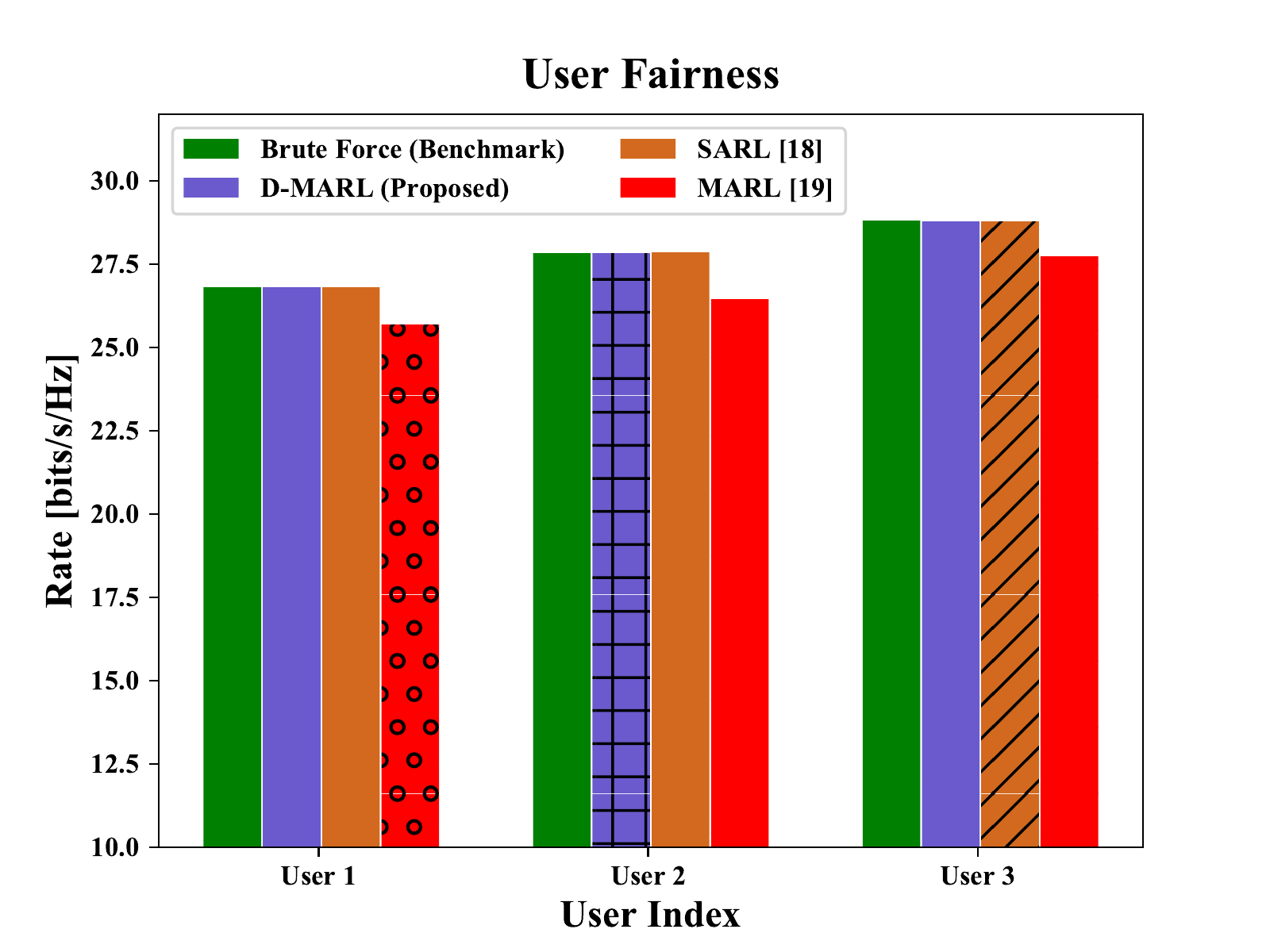} 
\caption{User fairness: AP coverage radius 250, SINR threshold $10$ dB} 
\label{rate_fairness}  
\end{figure}

Furthermore, a reliable and efficient network should ensure fairness while serving its associated users. 
A fair system delivers a nearly equal data rate to all users.  
Notice that our reward function, in Equation (\ref{RL_Reward}), designed such a way that it delivers a zero reward if any VU's downlink SINR is below the threshold level.
Therefore, the proposed solution guarantees a minimum data rate for all users.
Besides, since this threshold is equal for all VUs, each of them shall receive a nearly identical data rate.
From $250$ test episodes, user fairness - while conserving the maximized EE, is presented in Fig. \ref{rate_fairness}.
Our proposed D-MARL delivers a Jain's fairness index \cite{jain1999throughput} $\left(\frac{\left(\sum_{u_i\in \mathcal{U}} C_i(t)\right)^2}{|\mathcal{U}|\sum_{u_i\in \mathcal{U}C_i(t)^2}} \right)$ of $\approx 0.99915$.
The fairness index for the optimal scheme, SARL \cite{lin2016qos} and MARL \cite{8664589} are $0.99915$, $0.99915$, and $0.99899$, respectively. 
Note that this fairness index varies between $0$ and $1$. 
A fairness index of $0$ means there is no fairness whatsoever in the network. Fairness among users increases with the rise of this index.

\section{Conclusion}  
\label{Conclusion}
In this paper, we have jointly optimized virtual cell formation and power allocation to assure ubiquitous connectivity and reliability at the vehicular edge networks for connected transportation. Thanks to RL's powerful complex problem-solving ability, the hard combinatorial joint optimization problem is efficiently solved using this sophisticated learning process. 
While we exploited several RL models, we conclude that a distributed learning approach on the centralized action-space delivers better performance over the traditional collaborative MARL model.
Especially, our proposed sophisticated D-MARL solution attains near-optimal benchmark performance within a nominal number of training episodes.

\bibliography{Reference.bib}
\bibliographystyle{IEEEtran}
	
\end{document}